\documentstyle{elsart}
\newcommand{\be}{\begin{equation}}
\newcommand{\bef}{\begin{figure}}
\newcommand{\eef}{\end{figure}}
\newcommand{\hmp}{h^{-1}Mpc}
\newcommand{\etal}{{\em et al.}}
\newcommand{\ee}{\end{equation}}
\def\spose#1{\hbox to 0pt{#1\hss}}
\def\ltapprox{\mathrel{\spose{\lower 
3pt\hbox{$\mathchar"218$}}
 \raise 2.0pt\hbox{$\mathchar"13C$}}}
\def\gtapprox{\mathrel{\spose{\lower 
3pt\hbox{$\mathchar"218$}}
 \raise 2.0pt\hbox{$\mathchar"13E$}}}
\def\inapprox{\mathrel{\spose{\lower 
3pt\hbox{$\mathchar"218$}}
 \raise 2.0pt\hbox{$\mathchar"232$}}}

\begin{document}
\begin{frontmatter}
\title{Finite size effects on the galaxy number counts:
evidence for fractal behavior up to the deepest scale}
\author[Roma,Infm,Bologna]{F. Sylos Labini},
\author[Roma]{A. Gabrielli}, 
\author[Roma,Infm,Calabria]{M. Montuori} and
\author[Roma,Infm]{L. Pietronero}
\address[Roma]{Dipartimento di Fisica, Universit\`a di Roma
``La Sapienza'' P.le A. Moro 2, I-00185 Roma, Italy.} 
\address[Infm]{INFM, Sezione di Roma 1}
\address[Bologna]{Dipartimento di Fisica, Universit\`a
 di Bologna, Italy}
\address[Calabria]{Dipartimento di Fisica, Universit\`a
 della Calabria, Italy}

\begin{abstract}
We introduce and study two new concepts which are essential
for the quantitative analysis of the statistical quality of the available
galaxy samples. These are the dilution effect and the small scale
fluctuations. We show that the various data that are considered as pointing 
to a homogenous distribution are all affected by these spurious effects and
their interpretation should be completely changed. In particular, 
we show that finite size effects strongly affect 
the determination of 
the galaxy number counts, namely
the number versus magnitude relation ($N(<m)$)
as computed from the origin. 
When one computes $N(<m)$  averaged over 
all the points of a redshift survey 
one observes an exponent $\alpha = D/5 \approx 0.4$
compatible with the fractal dimension $D \approx 2$ 
derived from the full correlation analysis.
Instead the observation of an exponent $\alpha \approx 0.6$ 
at relatively small scales, where the distribution is certainly 
not homogeneous, is shown to be 
related to finite size effects. 
We conclude therefore that the observed counts correspond
to a fractal distribution with dimension $D \approx 2$
in the entire range $12 \ltapprox  m \ltapprox 28$,  
  that is to say the largest scales ever
 probed for luminous matter.
In addition our results 
permit to clarify various problems of the angular catalogs, and
to show their compatibility with the fractal behavior.
We consider also the distribution of Radio-galaxies, Quasars
and $\gamma$ ray burst, and we show their compatibility
with a fractal structure with $D \approx 1.6 \div 1.8$.
Finally we have established a quantitative criterion 
that allows us to
define and {\em predict} the statistical validity of a galaxy catalog
 (angular or three 
dimensional).
\end{abstract}
\end{frontmatter}

\section{Introduction}

The crucial question we are going to discuss in detail is the
 definition of the minimal statistics that is necessary to characterize
 correctly the large scale distribution of matter. The clarification
 and the definition of this fundamental point will allow us to
 reinterpret and clarify the conflictual situation arising from
 observations of different nature.

The best information on galaxy distribution would be to know the
 position in space of all the galaxies. The data that are closest to
this ideal situation are the so-called "volume-limited" (VL) samples
\cite{dp83},\cite{cp92}. The VL samples can be extracted from a
 redshift survey that contains information on the three spatial
 coordinates (angular position plus redshift). Such samples avoid
 the luminosity selection effect related to the observational point:
 one defines a maximum depth and includes in the sample only
 these galaxies that would be visible from any point of this volume.
 In such a way this VL sample is observer independent and its
 correlation properties can be studied directly. 

The galaxy distribution in all the redshift catalogues currently
 available \cite{kos81}-\cite{delapp95} show a
 highly inhomogeneous distribution characterized by the presence of
 clusters, superclusters, filaments walls and large voids up to the
 boundaries of such surveys.

The modern statistical analysis of these three dimensional samples
 \cite{pie87} 
show that this irregular distribution is 
characterized by having long range fractal correlations. In
 particular Coleman \& Pietronero \cite{cp92}, \cite{cp88} found
 that the galaxy distribution in the CfA1 survey \cite{hd83} is
 fractal with $D \approx 1.6$ up to $R_s \approx 20 \hmp$. This
 depth is of the order of the radius of the maximum sphere that is
 fully contained in the sample volume. This limit is imposed by the
 request that the correlation function should be computed only in
 {\em spherical shell}
(\cite{cp92}).  
 There have been attempts to push $R_s$ to larger values by using
 various
weighting schemes for the treatment of boundary conditions
 \cite{gu92}.
 These methods
however, unavoidably introduce artificial homogenization effects
 and 
therefore
should be avoided \cite{cp92}.
Sylos Labini {\em et al.} \cite{slmp95} found
 that the Perseus-Pisces galaxy redshift surveys \cite{hg88} has
 fractal long-range correlations up to $30 \hmp$ with dimension $D
 \approx 2$. The fractal dimension in this case is somewhat larger
 than in CfA1. This tendency for a higher value of the fractal
 dimension is related to the poor statistics and small depth of the
 CfA1 catalogue
and it is confirmed by various other redshift surveys.  
For example a similar behavior has been found by Di Nella \etal 
\cite{dmpps95} and \cite{ds95}
in the LEDA database \cite{pb88}
up to $\sim 150 h^{-1}Mpc$. 
 A first apparent contradictory result has been found by Strauss
 \etal \cite{str92} (see also \cite{fis94}) 
analyzing the IRAS catalogues,
 where they found an evidence towards homogenization at $R_s
\approx 15 \div 20 \hmp$.  

A different way to get information for larger scales is to
compute the conditional average (in VL samples) from the 
observation point
only. This allows us to extend $R_s$ to the full depth of the 
catalogue, at the
expenses of having a reduced statistics. We are going to see that in 
this way it
is possible to detect fractal correlations up to $130 \hmp$ in 
Perseus-Pisces (PP) \cite{slmp95} and up to
$800 \div 900\hmp$ in the new ESP catalogue \cite{bslmp94}.

Historically \cite{hu26} \cite{pe93} 
the oldest type of data about galaxy distribution is given by
the relation between the 
number of observed galaxies $N(>f)$  and their apparent
brightness $f$. It is easy to show that \cite{pe93}
\begin{equation}
\label{eq1}
N( >f)  \sim  f^{-\frac{D}{2}}
\end{equation}
where $D$
 is the fractal dimension of the galaxy distribution. Usually this
relation is written in terms of the apparent magnitude 
$f \sim 10^{-0.4 m}$ 
(note that bright galaxies correspond to small $m$).
In terms of $m$, Eq.\ref{eq1} becomes
\be
\label{nn1}
\log N(<m)   \sim \alpha m
\ee
 with $\alpha = D/5$ 
\cite{bslmp94} \cite{pe93}. 
Note that $\alpha$ is the coefficient of the exponential 
behavior of Eq.\ref{nn1} and we will call it "exponent" even though
 it  should not be confused with the exponents of power 
law behaviors.
In Fig.\ref{f1}
\bef
\vspace{5cm}
\caption{
\label{f1}
The galaxy number counts in the $B$-band,
from several surveys.
 In the range $12 \ltapprox  m \ltapprox 19$ 
the counts show an exponent $\alpha \approx 0.6$, while 
in the range $19 \ltapprox m \ltapprox 28$ 
the exponent is $\alpha  \approx 0.4$.
}
\eef
we have collected all the
recent observations of $N(<m)$ 
versus $m$ \cite{ss84}-\cite{ms91}
in the $B$-spectral-band ($m_B$).
At bright and intermediate magnitude 
($\:12 \ltapprox m_B \ltapprox 18$), 
corresponding to small redshift
($\:z<0.2$),  one obtains $\:\alpha \approx 0.6$, while  
from $m_B \sim 19$ up to $m_B \sim 28$ 
the counts are well fitted by a smaller exponent with 
$\alpha \approx  0.4$.  The usual interpretation
\cite{pe93}, \cite{yt88}-\cite{mg94}  is that $\alpha \approx  0.6$
corresponds to $D \approx 3$ consistent with homogeneity, while at 
large scales
galaxy evolution and space time expansion effects are invoked to
 explain 
the lower value
$\alpha  \approx 0.4$.
On the basis of the previous discussion of the VL samples we can 
see that this
interpretation is untenable. In fact, we know for sure that, at least
 up 
to $R_s \sim 150  \hmp $
there are fractal correlations \cite{cp92}, \cite{slmp95}, \cite{dmpps95}
 so one would eventually expect the
opposite behaviour. Namely  small value of 
$\alpha \approx 0.4$ (consistent with $D \approx 2$) at
small scales followed by a crossover to an eventual homogeneous 
distribution at
large scales ($\alpha  \approx 0.6$ and $D  \approx 3$).

The GNC in the (red) $R$-band shows an exponent $\alpha\approx
 0.37-0.41$
in the range $20<R<23$ 
\cite{ty88},\cite{jf91} and \cite{ms91}.
Moreover Gardner {\em et al.} \cite{gc93}  have studied the GNC
in the (infrared) 
{\em K}-band in the range $\:12 \ltapprox K \ltapprox 23$, and
they show that the slope of the counts changes at 
$K \approx 17$ 
from $\:0.67$ to $\:0.26$ 
(see also \cite{co94}-\cite{so94}).
Djorgovski {\em et al.} \cite{dj95} found 
that the slope of the GNC is  
little bit higher than \cite{gc93}, i.e.  $\alpha =0.315 \pm 0.02$ 
between $K=20$ and $24$ magn.
The situation is therefore 
quite similar in the different spectral bands.
The puzzling behaviour of the GNC represents the {\em second}
contradiction we find in the data analysis.

An additional argument in favour of homogeneity at rather small
 scales has been proposed as arising from an appropriate rescaling
 of the angular correlations \cite{pe93}. This is the {\em third}
 evidence that seems to be conflictual with the properties
 observed in the VL correlations analysis.

We argue here  that this  apparently 
contradictory experimental situation can be 
fully understood on  the light 
of the small scale effects in the 
space distribution of galaxies. For example a
fractal distribution is non analytic in every 
occupied point: it is not possible to
 define a meaningful average density 
because  we are dealing with intrinsic 
fluctuations that grow with  as the scale of the system itself.
This situation is qualitatively different from an homogeneous
 picture, in which a well defined density exists, and the fluctuations 
represent only  small amplitude perturbations. The nature of the 
fluctuations in these two cases is completely different, and for 
fractals the fluctuations themselves define all the statistical
 properties of the distribution.
This concept has  
 dramatic consequences in the following discussion as well as  
in the  determination of various observable quantities, 
such as the amplitude of the two point
 angular  correlation function.

It is worth to notice that small scale effects are usually neglected
in the study of fractal structures because one can generate large 
enough structures to avoid these problems. In Astrophysics the data 
are instead intrinsically limited and we are going to see that an 
analysis of finite size effects is very important
We discuss in detail the problems of finite size effects
in the determination of the asymptotic properties of fractal 
distributions,  considering explicitly the problems 
induced by the lower cut-off
(Section 3).
  
We discuss in detail the case of real galaxy redshift survey 
(Sec.4 and Sec.5), while in Section 6 we consider 
  the case of the 
{\em magnitude limited} samples. 
In particular, we find  a criterion which allows us to 
define the statistical validity of the various samples
with  respect to the finite size effects.
In particular it results that the apparent exponent 
$\alpha \approx 0.6$  at small scales 
(bright magnitudes $m<19$) 
arises purely from finite size   effects
and cannot be related to the real correlation properties
of the sample. 

In Section 7 we discuss the concept of a {\em statistically
fair sample}, i.e. a sample which contains enough
statistical information to be representative of the whole
distribution from which it has been extracted. 
In particular we define  the lower number density of points
needed to recover the genuine statistical properties of a certain
distribution (homogeneous or fractal).
 Moreover we find that the all-sky IRAS surveys \cite{str95}, 
\cite{str92}
do not contain enough points to be {\em statistically fair} samples.

The implications of our analysis 
is that it provides a quantitative criterion to define the statistical
 validity
of {\em a redshift and an angular survey}, and for the optimization 
of its geometry to obtain the maximum reliable 
information (Sec.8). These results can be useful also
to design future programs for redshift or angular surveys. 
On this basis we can {\em predict} 
the expected statistical properties of several surveys
such as CfA2 \cite{gh89}, Las Campanas \cite{sc92},
ESP \cite{ve94}, etc.

In Section 9 we discuss  the 
angular correlation functions, 
on the light of the previous analysis.
It results that the scaling of the amplitude of 
the angular correlation function
with the depth of the sample, that is considered
to be a possible evidence for of homogeneity 
\cite{pe93},
is actually due to   finite size effects just as the 
exponent of the number counts.

In addition in Section 10 we consider the distribution of
Radio-galaxies, Quasars and $\gamma$ ray bursts. We show
that these objects are fractally distributed with 
$ D \approx 1.6 \div 1.8$.

Finally in Section 11 we present our main conclusion,
stressing in particular the dramatic and puzzling consequences 
on the general theoretical framework and, in 
particular on the Big-Bang model.

\section{Galaxy number counts: Basic Relations}
The basic assumption that we use to  compute all the following
 relations 
is that:
\be
\label{p1}
\nu(L,\vec{r}) = \phi(L) D(\vec{r})
\ee
i.e. that the number of galaxies for unit luminosity
and volume $\nu(L,\vec{r})$ can be separated as 
the product of the space density
$ D(\vec{r})$  and  the luminosity function $ \phi(L)$.
This is a crude approximation 
in view of the multifractal properties of the distribution 
(correlation between position and luminosity), and a detailed 
discussion is found in  \cite{slp95}
and \cite{bslmp94}.
However, for the propose of the present discussion
the approximation of Eq.\ref{p1} is rather good and the 
explicit consideration of the multifractal properties 
would have a minor effect on the properties we are going to 
discuss (\cite{slp95}).

The integrated space number density, .i.e. 
the total number of points inside a sphere of radius $R$, 
 in the general case,  has the following property
\be
\label{p2}
N(<R) = \int_{0}^{R} D(\vec{r}) d^3 r = 
\int_{0}^{R} \sum_{i=1}^{N} \delta(\vec{r}-\vec{r_i}) d^3 r = BR^D
\ee
where $D=3$ for the homogeneous case, 
while if $D<3$ the distribution is fractal
and $B$ is the prefactor (see Sec. 3).
This is an asymptotic relation and the problem is to recover 
this behavior from finite size portions of real structures
(we discuss in detail this point in Section 3).
 
We briefly introduce some basic definitions.
If $L$ is the absolute or intrinsic 
luminosity of a galaxy at distance $r$,
this will  appears with an apparent flux
\be
\label{l1}
f = \frac{L}{4 \pi r^2}.
\ee
Suppose all the galaxies have the same luminosity $L$.
If the number of points inside a sphere of radius $r$ grows 
as $N(<r) \sim r^D$, then it is simple to show that the number
of galaxies with apparent flux greater than $f$ 
goes as $N(>f) \sim f^{-\frac{D}{2}}$. This the basic
relation for the GNC. Now we consider the case in which
there is a certain distribution of galaxy luminosities.

For historical reasons the apparent magnitude $m$ 
of an object with incoming flux $f$ 
is defined to be (\cite{pe93})
\be
\label{l2}
m = -2.5 \log_{10}f + constant,
\ee
while the absolute magnitude $M$
 is instead related to its intrinsic luminosity $L$ by
\be
\label{l3}
M=-2.5 \log_{10}L + constant.
\ee
From Eq.\ref{l1} that 
the difference between the apparent and the absolute magnitudes
 of an
object at distance $r$ is (at relatively small distances, 
neglecting relativistic effects)
\be
\label{mm}
m-M = 5 \log_{10}r + 25
\ee
where $r$ is expressed in Megaparsec ($1 Mpc= 3.26 \cdot 10^{6}
 light \; years.$)

We now compute the expected GNC in 
the simplest case of a magnitude limited (ML) 
sample. 
A ML sample is obtained measuring all the galaxies with 
apparent magnitude brighter than a certain limit $m_{lim}$.
In this case we have (for $m < m_{lim}$)
\be
\label{q3}
N(<m) = 
B \Phi(\infty) 10^{\frac{D}{5}m}
\ee
where
\be
\label{q3a}
\Phi(\infty) = \int_{-\infty}^{\infty} 
\phi(M) 10^{-\frac{D}{5}(M+25)} dM 
\ee

We consider now the case of a volume limited (VL) sample.
A  VL  sample contains every galaxy in the volume
which is more luminous than a certain limit, so that in such a
 sample
there is no incompleteness for an observational 
luminosity selection effect (\cite{dp83}, \cite{cp92}).
Such a sample is defined by a certain maximum distance $R$ 
and an absolute magnitude limit given b:
\be
\label{q3b}
M_{lim}=m_{lim}-5log_{10}R -25
\ee
($m_{lim}$ is the survey apparent magnitude limit).
By performing the calculations 
for the number-magnitude relation, 
  we obtain 
\be
\label{q4b}
N(<m)= A(m) \cdot 10^{\frac{D}{5}m} + C(m)
\ee
where $A(m)$ is
\be
\label{q4c}
A(m) =B \int_{M(m)}^{M_{lim}}\phi(M) 10^{-\frac{D}{5}(M+25)}dM 
\ee
and $M(m)$ is given by Eq.\ref{mm} (with $R_{VL}$ in the place of $r$),
 and it is a function of $m$. 
The second term is
\be
\label{q5e}
C(m)= BR^{D}\int_{-\infty}^{M(m)} \phi(M) dM
\ee
This term, as $A(m)$,
 depends from the VL sample considered. 
We assume a luminosity function 
with a Schechter shape \cite{sc76}
\be
\label{q5f}
\phi(M)dM \sim 10^{-0.4(\delta+1)M}e^{-10^{0.4(M^*-M)}}dM
\ee
where $\delta \sim -1.1$ and the cut-off $M^*\sim -19.5$ 
\cite{dac94}.
For $M(m) \gtapprox M^{*}$ we have that $C(m) \approx 0$,
and $A(m)$ is nearly constant with $m$.
This happens in particular for the deeper VL samples
for which $M_{lim}\sim M^*$.
For the less deeper VL ($M_{lim} > M^*$) samples these terms 
can be considered as a deviation from a power law
behavior only for
$m \rightarrow m_{lim}$.
In Fig.\ref{f2}
\bef
\vspace{5cm}
\caption{\label{f2}
 Behaviour of $N(<m)$ for various VL samples
of an ideal survey with $m_{lim}=15.5$.
For the deeper VL sample ($120 h^{-1}Mpc$)  (large squares) 
a straight power law is found, 
while for the $40 h^{-1}Mpc$  (small squares) 
sample there is a flattening for $m \rightarrow
m_{lim}$, even if the true exponent 
$\alpha =D/5$ is found at lower magnitudes. 
The sample at $80 h^{-1}Mpc$  (crosses)
show an intermediate situation.
}
\eef
we show the behavior of Eq.\ref{q4b} for the case 
$m_{lim}=15.5$ and we consider VL samples at 
VL40: ($R=40 h^{-1}Mpc, M^*=-17.53 $); VL80 ($R=80 h^{-1}Mpc,
 M^*=-19.07 $);
VL120: ($R=120h^{-1}Mpc,M^*=-19.93$).
We can see that the first subsample for $m \rightarrow m_{lim}$
show a  curved behavior while in the other cases 
the $\log(N(<m))-m$ relation is well fitted by a power law
with the right exponent $\alpha=D/5$. This is
the ideal case where no finite size effects are present.

If one has $\phi(M) = \delta(M-M_0)$ then it is simple to show that
$\log(N(<m)) \sim (D/5)m$ also each  VL sample. This 
relation will be useful in the following (Section 5).

\section{The problem of finite size effects}

In this section we discuss the general problem of the minimal 
sample size that is able to provide us with a statistically
 meaningful information.
For example the mass-length relation for a   fractal,
that defines the fractal dimension is (\cite{man82} and \cite{sie89})
\be
\label{e1}
D=\lim_{r \rightarrow \infty} \frac{\log(N(<r))}{\log(r)}
\ee
 However this relation is
 {\em properly defined only in the
asymptotic limit}, because only in
 this limit
 the fluctuations of the fractal structure are
 self-averaging.
 A fractal distribution is characterized by 
large fluctuations at all scales
 and these fluctuations determine all the
statistical properties of the structure.
If the structure has a lower cut-off,
as it is the case for any  real fractal, 
one needs a {\em "very large sample"} in order to recover 
the statistical properties of the
distribution itself.
Indeed, in any real physical
problem  we would like to recover the asymptotic properties
from the knowledge of a  {\em finite portion}
of a fractal  
and the problem is that   a single finite realization of a random
fractal is affected 
by finite size fluctuations due to the  lower cut-off.

Considering  homogeneous distribution we can define, an
 average, a characteristics 
volume associated to each particle. 
This is the Voronoi volume \cite{vo08} 
$v_v$ whose radius $\ell_v$ is of the order of the mean particle
 separation.
It is clear that the statistical 
properties of the system can be defined only in volumes
much larger than $v_v$. Up to this volume
in fact we observe essentially nothing.
Then one begins to include a few (strongly fluctuating)
points, and finally, the correct scaling behavior is recovered
 (Fig.\ref{f3}). 
\bef
\vspace{5cm}
\caption{\label{f3}
 Behavior of the density computed from one point.
At small distances, inside the Voronoi's length $V$,
one finds  almost no galaxies
because the total number is rather small.
Then the number of galaxies starts
to grow, but this regime is strongly affected by
finite size fluctuations.
Finally the correct scaling region $r \approx \lambda$
is reached.
In the intermediate region
the density can be approximated
roughly by a constant value.
  This leads to an apparent exponent $D \approx 3$ (in terms of GNC
$\alpha  \approx
0.6$).  This exponent is not real but just due to the
size effects.
}
\eef
For a Poisson
sample consisting of $N$ particles inside a 
volume $V$ then the
Voronoi volume (Fig.\ref{f4}) 
\bef
\vspace{5cm}
\caption{\label{f4}
Definition of the Voronoi volume (see text) 
}
\eef
is of the order
\be
\label{v1}
v_v \sim \frac{V}{N} 
\ee
and $\ell_v \approx v_v^{1/3}$.
In the case of homogeneous distribution, where 
the fluctuations have  {\em small amplitude}
with respect to the average density, one  readily recovers
the statistical properties of the system at small distances, say,
$ r \gtapprox 5 \ell_v$.

The case of fractal distribution is more subtle.
For a self-similar distribution one has, within a 
certain radius $r_0$, $N_0$ objects. Following \cite{cp92}
we can write the mass-length relation between 
$N(<R)$, the number of points inside a sphere of radius $R$,
and the distance $R$ of the type
\be
\label{new1}
N(<R) = B R^D 
\ee
where the prefactor $B$ is related to the lower cut-offs $N_0$ and
$r_0$
\be
\label{new2}
B=\frac{N_0}{r_0^D}
\ee
In this case the prefactor $B$ is defined for 
spherical  samples. If we have a portion of a sphere characterized
by a solid angle $\Omega$, we write Eq.\ref{new1} as
\be
\label{new1a}
N(<R) = B R^D \frac{\Omega}{4\pi}
\ee 
Suppose we have only a finite portion of a fractal structure
 characterized by 
a volume
\be
\label{new3}
V(R) = \frac{\Omega R^3}{3}
\ee
In this case the Voronoi's  volume can be 
written as 
\be
\label{new4}
v_v = \frac{4\pi}{3} \frac{R^{3-D}}{B}
\ee
We define 
\be
\label{new4b}
V_{\Omega}= \frac{\lambda^3}{3} \Omega 
\ee
as the minimal volume which allows 
us to recover the statistical properties of the system.
 From a series of numerical tests on artificial 
distributions we will conclude that $V_{\Omega}\sim 10^{3} v_v$.
 From this we can define the 
{\em minimal statistical length} as
\be
\label{new5}
\lambda \sim 
\left( \frac{4\pi}{\Omega} 10^3 \frac{R^{3-D}}{B} 
\right)^{\frac{1}{3}}
\ee
The {\em minimal statistical length} $\lambda$ is 
an explicit function of the prefactor $B$ and of
the depth of the survey $R$.
Essentially this discussion implies that $\lambda$ should 
be about one order of magnitude larger than $\ell_v \sim
 v_v^{1/3}$,
apart from solid angle effects.
We note that in order to optimize the 
region of the sample beyond the {\em minimal statistical length},
we have to maximize the ratio
\be
\label{new5a}
\frac{R}{\lambda} \sim R^{\frac{D}{3}} \Omega^{\frac{1}{3}}
\ee
hence if $D > 1$ it is clearly 
convenient to increase the depth of the 
 sample rather than the 
solid angle in order to improve the statistics 
as we discuss in detail in the following
(see in particular Sec.7).

In the case of real galaxy catalogs we have to consider the 
luminosity selection effects. In a volume 
limited (VL) sample, characterized by an absolute magnitude limit
$M_{lim}$, (minimal absolute flux) the mass-length relation
 Eq.\ref{new1a}, 
 can be generalized as (see Sec. 2)
\be 
\label{new6}
N(R,M_{lim}) = B R^D \frac{\Omega}{4 \pi} \psi(M_{lim})
\ee
where $\psi(M_{lim})$ is the probability that a galaxy has 
an absolute magnitude brighter than $M_{lim}$
\be
\label{new7}
0 < \psi(M_{lim})  = \frac{\int_{-\infty}^{M_{lim}} \phi(M) dM}
{\Psi(\infty)} < 1
\ee
where $\phi(M)$ is the Schechter luminosity function (Sec.2) 
and $\Psi(\infty)$ is the normalizing factor
\be
\label{new8}
\Psi(\infty) = \int_{-\infty}^{M_{min}} \phi(M) dM
\ee
where $M_{min}$ is the fainter absolute magnitude surveyed 
in the catalog (usually $M_{min} \approx -10 \div -11$).

It is possible to compute the intrinsic prefactor $B$ 
from the knowledge of the 
average conditional density
$\Gamma(r)$ \cite{cp92}, \cite{slmp95} as 
\be
\label{new9}
\Gamma(r) = \frac{D}{4\pi} B r^{D-3}
\ee
computed in the VL samples and normalized for the 
luminosity factor (Eq.\ref{new7}). 
In the various VL  subsamples
of Perseus-Pisces and CfA1 redshift surveys we find that
\be
\label{new10}
B \approx 10 \div 15 (\hmp)^{-D}
\ee
depending on the parameters of the Schechter function $M^*$ 
and $\delta$.
From Eq.\ref{new5} and Eq.\ref{new10}  we obtain
\be
\label{new11}
\lambda \approx \left( \frac{10^3 R^{3-D}}{\Omega}
 \right)^{\frac{1}{3}}
\ee
and for $R$ in the range $100 \div1000 \hmp $ we find that 
a good approximation to Eq.\ref{new11} is
\be
\label{v3}
\lambda \approx \frac{(40 \div 60) \hmp}{\Omega^{\frac{1}{3}}}
\ee
 This is the value of the {\em minimal statistical length} that
we will use in the following (see Table 1).
\begin{table}
\caption{The {\em minimal statistical length}
$\lambda$ for several redshift surveys }
\begin{tabular}{|c|c|c|c|}
\hline
	&	&	&\\
\rm{Survey} & $\Omega (sr)$ & $\lambda (\hmp)$  & References \\
&	&	&\\
\hline
\hline
	&	&	&\\
CfA1 & 1.8 	& 30 & Huchra {\em et al.}, (1983); Davis {\em et al.}, 
(1982)  \\
	&	&	&\\
CfA2 & 1.8 	& 30& Geller \& Huchra, (1988); Park {\em et al.}, (1994)  \\
	&	& 	&\\
SSRS1 & 1.13 	& 40  & Da Costa {\em et al.}, (1988; 1991) \\
	&	& 	&\\
SSRS2 & 1.13 	& 40  & Da Costa {\em et al.}, (1994)\\
	&	& 	&\\
PP & 1 	& 50  & Haynes {\em et al.}, (1988); Sylos Labini {\em et al.}, (1995)\\
	&	& 	&\\
LEDA    & 4 $\pi$ 	& 20 & Paturel {\em et al.}, (1988)\\
	&	& 	&\\
ORS & 8 	& 25    &  Santiago {\em et al.}, (1994)\\
	&	& 	&\\
APM & 1.36 	& 50& Loveday {\em et al.}, (1992a, b) \\
	&	& 	&\\
LCRS & 0.12 	& 100  & Schectman, (1992)\\
	&	& 	&\\
KOSS & 0.01 	& 250  & Kirshner {\em et al.}, (1981)\\
	&	& 	&\\
ESP & 0.006 	& 300  & Vettolani {\em et al.}, (1994)\\
	&	& 	&\\
KOO & $1.5 \cdot 10^{-4}$ 	& 950  & Koo, (1988)\\
	&	& 	&\\
\hline
\end{tabular}
\end{table} 
This length depends also, but weakly, from the particular
morphological features of the sample.
In the case of Perseus-Pisces ($\Omega = 0.9$) 
we find $\lambda \approx 50 \hmp $
as shown in Fig.\ref{f5} and Fig.\ref{f6}, 
\bef
\vspace{5cm}
\caption{\label{f5}
The spatial density $n(r)$ and $N(<r)$ 
computed in two VL sample
cut at $60 h^{-1}Mpc$ ($a,b$), and $70 h^{-1} Mpc$ ($c,d$). 
The density is dominated by large
fluctuations and it has not reached the scaling regime.
}
\eef
\bef
\vspace{5cm}
\caption{\label{f6}
The spatial density $n(r)$ and $N(<r)$ computed in two VL sample
VL110 ($a,b$) and VL130 ($c,d$).
 In this case the density is dominated by large
fluctuations only at small distances, 
while at larger distances, after the Perseus Pisces chain at $50 h^{-1}Mpc$, 
a very  well defined power law behavior is shown, with the same
exponent of  the correlation function of Fig.8 ($D=2$)
}
\eef
and for CfA1 
 ($\Omega =1.8$)  $\lambda \approx 30 \hmp $ (Fig.\ref{f7}):
\bef
\vspace{5cm}
\caption{\label{f7}
The spatial density $(a)$ $n(r)$ and the 
integral number of points $(b)$ $N(<r)$ computed in a VL sample
(VL60) of CfA1. 
A very  well defined power law behavior with $D \approx 2$
 is shown for $r \gtapprox 30 \hmp \approx \lambda$.
}
\eef
in both cases the agreement with Eq.\ref{v3} is quite good.
Moreover we have done the following test: we have cut the 
Perseus-Pisces survey at various solid angles and we 
have checked that Eq.\ref{new11} holds 
with good accuracy for various values of $R$ and $\Omega$.

For relatively small volumes it is possible
to recover the correct scaling behavior for scales of 
order of $\ell_v$ (instead of $ \sim 10 \ell_v$) 
by averaging over several samples or,
as it happens in real cases, over several points of the 
same sample when this is possible. 
Indeed when we compute 
the correlation function we perform an average over all
the points of the system even if the VL  
sample is not deep enough to satisfy the condition 
expressed by Eq.\ref{v3}. Even in this case the lower cut-off
 introduces 
a limit in the sample statistics as we point out in 
Sec.7 and Sec.8.
 
In Section 4 we show how these considerations can be applied to 
 the 
case of a real redshift survey (Perseus-Pisces) as
well as in the case of artificial catalogs with a 
priori assigned properties.   

\section{Integral from the vertex in Perseus-Pisces}

To clarify the effects of the spatial inhomogeneities 
and finite size effects 
we have studied the GNC in the Perseus-Pisces redshift
 survey \cite{hg88}.
In a previous paper (\cite{slmp95}) we have analyzed  the spatial properties 
of galaxy distribution in this sample 
and we briefly summarize our main results. We find that 
the correlation 
function $\Gamma(r)$ (conditional average density - \cite{cp92}), 
computed in several  VL samples, is  
\be
\label{g1}
\Gamma(r) = \frac{<n(\vec{r_0})n(\vec{r}+\vec{r_0})>}{<n>} 
\sim r^{-\gamma}
\ee
where $\gamma \approx 1$ up to 
$\sim 30 h^{-1}Mpc$. This result means 
that the galaxy distribution in this 
sample is fractal up to this depth 
with dimension $D = 3 -\gamma  \approx 2$.
This trend of larger value of $D$ 
with respect to CfA1 ($D \approx 1.5$) 
 is confirmed by all the deeper samples
(CfA2 \cite{pv94} \cite{sla95}, LEDA \cite{dmpps95} \cite{ds95},
 ESP \cite{psl94})
and it is probably due to a more stable 
statistics (see Sec.7) of this larger sample with 
respect to CfA1.
We have also studied the behavior of
 the galaxy number density in the
VL samples, i.e. 
the behavior of:
\be
\label{g2}
n(r) = \frac{N(<r)}{V(r)} \sim r^{D-3}
\ee
  One expects that, if 
the distribution is homogeneous the density is  
 constant,  while if it is fractal it decays 
with a power low behavior.

When 
one computes 
the correlation 
function of Eq.\ref{g1}, 
one indeed performs an average over all the points of the survey.
In particular we limit our 
analysis to the size defined by the 
radius of 
the maximum sphere fully contained 
in the sample volume, so that we do not
introduce any weighting scheme in the 
treatment of the boundaries of the sample \cite{slmp95}.
Several authors (e.g. \cite{gu92}) have extended the effective depth 
for the computation of $\Gamma(r)$ making use 
of the weighting schemes for the treatment of the boundary
 conditions:
Guzzo \etal \cite{gu92} found a tendency towards homogenization at 
$\sim 40 \hmp$ in PP, while we have limited our analysis at $R_s
 \approx 30 \hmp$
As we show in the following, such a method introduces a spurious 
tendency towards homogenization, and must be avoided for
 system characterized by long-range correlation (see also
 \cite{cp92} and \cite{slmp95}). 

On the contrary Eq.\ref{g2} is computed only from
a single 
 point, the origin.
This allows us to extend the study of the spatial 
distribution up to very deep scales and in fact
 we find that Eq.\ref{g2} holds up to 
$\sim 130 h^{-1}Mpc$, that is the maximum 
depth surveyed by this catalog,
with the same exponent $\gamma \approx 1$ as before.
The price to pay is that this method is 
strongly affected by  statistical 
fluctuations and finite size effects.
Analogously,   
when one computes $N(<m)$,  one does not perform an average 
but  just counts the points from the origin. 
As in the case of $n(r)$ also $N(<m)$ will be strongly affected
by statistical fluctuations
 due to finite size effects. We now clarify how the
behaviour of $N(<m)$, and in particular its exponent,
 are influenced  by these effects.

We have computed the $n(r)$ in the various VL sample,
and we show the results   in Fig.\ref{f5} and Fig.\ref{f6}.
In the less deeper 
VL samples (VL60, VL70)  (Fig.\ref{f5}) 
we see that the density does not show 
any smooth behavior because in this case
the finite size effects dominate the behaviour as we are 
at distances $r < \lambda $ (Eq.\ref{v3}), while at about
 the same scales we can find 
a very well defined power law behavior 
with the correlation
 function analysis (Fig.\ref{f8}). 
\bef
\vspace{5cm}
\caption{\label{f8}
The correlation function for VL60 (small squares)
 and VL70 
crosses) and  VL110 (large squares).
 In this case the correlation function $\Gamma(r)$, that 
is the conditional density averaged from each point of the sample, 
shows a very well defined
power law behaviour. The finite size effects are eliminated  
 exactly for the averaging procedure. The reference line has a slope of 
$-\gamma=D-3=-1$
}
\eef

In the deeper VL samples (VL110, VL130) we can see 
(Fig.\ref{f6}) that a smooth behavior is reached for distances 
larger than
the scaling distance 
$r \approx \lambda \sim 50 h^{-1}Mpc$. The fractal dimension 
turns out to be $D \approx 2$ as 
in the case of the correlation
 function (Fig.\ref{f8}). Also the 
amplitude of the density matches  quite well that of $\Gamma(r)$
(if properly normalized).

We show in Fig.\ref{f9},
\bef
\vspace{5cm}
\caption{\label{f9}
The Number counts $N(<m)$ 
for the VL samples VL60 and VL70. The
 slope is $\alpha \approx 0.6$. In this case occurs a 
flattening for $m \rightarrow m_{lim}$.
}
\eef 
Fig.\ref{f10} 
\bef
\vspace{5cm}
\caption{\label{f10}
The Number counts $N(<m)$ for the VL samples
VL110 and VL130. The slope is $\alpha \approx 0.4$, a part
from the initial fast growth. 
This behavior correspond to a well defined
define power law behavior of the density with exponent 
$D \approx  5 \alpha \approx 2$ (Fig.6).
}
\eef
and Fig.\ref{f11}
\bef
\vspace{5cm}
\caption{\label{f11}
The Number counts $N(<m)$ for the whole 
magnitude limit sample.
The slope is $\alpha \approx 0.6$ and it is clearly 
associated only to
fluctuations in the spatial distribution rather 
than to a real 
homogeneity in space.
}
\eef the behaviour of
$N(<m)$ respectively 
for the VL samples of Fig.\ref{f5} (VL60 and VL70), 
of Fig.\ref{f6} (VL110, VL130) and finally for the whole 
magnitude limit sample. We can see that for VL60 and VL70 there
are very strong inhomogeneities in the behaviour of 
$n(r)$ and these are associated with a slope $\alpha \approx 0.6$
for the GNC. (The flattening for 
$m \rightarrow m_{lim}$ is just 
to a luminosity selection effect
 that is  explained in Section 2).
 For VL110 and VL130 the behaviour of the density is 
much more regular and smooth, so that it shows
 indeed a clear power law behavior.
Correspondingly the behaviour of $N(<m)$ is well fitted by 
 $\alpha \approx 0.4$. Finally the whole magnitude limit sample 
is again described by an exponent $\alpha \approx 0.6$.

We have now enough elements to describe the behaviour of the
 GNC.
The first point is that the exponent of the 
GNC is strongly related
to the space distribution. Indeed 
what has never been taken into account   before is
 the role of finite size effects.
The behaviour of the GNC is due to a convolution of the 
space density and the luminosity function (Sec.2), and
the space 
density enters in the GNC as an integrated quantity.  The
problem is to consider the right space 
density in the data analysis.
In fact,  
if we have a very fluctuating behaviour for the 
density in a certain region, as in  the case shown in Fig.\ref{f5},
its integral  over this range of length scales 
is almost equivalent to a flat one.
This can be seen also in Fig.\ref{f3}:
 at small distances
one finds  almost no galaxies 
because the total number is rather small.
Then the number of galaxies starts 
to grow, but this regime is strongly affected by  
finite size fluctuations.
Finally the correct scaling region $r \approx \lambda$ 
is reached.
This means  
for example that if one has a fractal distribution, there will be first a raise
of the density, due to finite size effects and 
characterized by strong  fluctuations, because no galaxies are 
present before a certain characteristic scale. Once 
one enters in the correct scaling regime for 
a fractal the 
 density becomes to decay as a power law.
So in this regime of raise and fall 
with strong fluctuations there will be a 
region in which  the density can be approximated roughly by a
 constant 
value.  This leads to an apparent exponent $D \approx 3$, so that 
the integrated 
number grows as 
  $N(<r) \sim r^3$ and  it is associated in terms of GNC, to
$\alpha  \approx 
0.6$.  This exponent is therefore not real but just due to the 
finite size fluctuations.
Only when a well defined
statistical scaling regime has been 
reached, i.e. for $r > \lambda$,
 one can find the genuine scaling properties of the 
  structure, otherwise the behaviour is
completely  dominated by spurious finite size effects.
 
The question of the difference between the
integration from the origin and correlation properties
 averaged over all points lead us to a subtle 
problem of {\em asymmetric fluctuations} in a fractal 
structure. From our discussion, exemplified 
by Fig.\ref{f3}, the region between $\ell_v$ and $\lambda$
 corresponds to an underdensity with respect to 
the real one.
However we have also showed that for the full 
correlation averaged over all the points ($\Gamma(r)$) 
the correct scaling is recovered at distances 
appreciably smaller than $\lambda$.
This means that in some points one 
should observe an overdensity between 
$\ell_v$ and $\lambda$.
However, given the intrinsic asymmetry 
between filled and empty regions in a fractal, 
only very few points will show 
the overdensity ( a fractal 
structure is asymptotically 
dominated by voids). These 
few points nevertheless will 
have an important effect 
on the average values of the correlations. 
This means that, in practice, a typical points 
shows an underdensity up to
$\lambda$ as shown in Fig.\ref{f3}.
The full averages instead converge 
at much shorter distances. This 
discussion shows the peculiar 
and asymmetric nature of finite 
size fluctuations in fractals 
as compared to the symmetric Poissonian case.

For homogeneous distribution (Fig.\ref{f12}) 
\bef
\vspace{5cm}
\caption{
\label{f12}
$(a)$ $n(r)$;  $(b)$ $N(<r)$ and $(c)$ $N(<m)$ 
(single point (squares), averaged (crosses)) for an 
homogeneous distribution. The Voronoi length is $\ell_v \approx 2 Mpc$.
In this case the finite size effects do  not  
affect too much the properties of the system 
because an homogeneous distribution is characterized by
{\em small amplitude fluctuations}. This situation is qualitatively
different from the fractal one, and 
 for distance 
$r \gtapprox (2-3) \ell_v$ the correct
scaling regime 
is readily found.
}
\eef
the situation is in fact quite different. Below 
the Voronoi length $\ell_v$ there are 
finite size fluctuations, but for distances 
$r \gtapprox (2\div 4) \ell_v$ the correct
scaling regime is readily found for the density, the integrated
density and the number counts. 
In this case the finite size effects do not 
affect too much the properties of the system 
because a Poisson distribution is characterized by
{\em small amplitude fluctuations}. 

In the VL samples where $n(r)$ scales with the asymptotic properties 
(Fig.\ref{f6}) the GNC grows also with the right exponent ($\alpha=D/5$).
If we now consider instead the behaviour of the GNC in the whole
magnitude limit survey, we fund that 
the exponent is $\alpha  \approx 0.6$ (Fig.\ref{f11}).
 This behavior can be understood by considering that at small distances, 
well inside the distance $\lambda$ defined by Eq.\ref{v3}, 
the number
of galaxies present in  the sample 
is large because there are  
 galaxies of all magnitudes. 
Hence the majority of 
galaxies correspond to small distances ($r < \lambda$) 
and the distribution has not
reached the scaling regime  in 
which the  statistical self-averaging 
  properties of the system are present. For this reason in the ML 
sample the finite size 
fluctuations dominate completely the behavior of the GNC.
Therefore this 
behaviour  in the ML sample is 
associated with spurious finite 
size effects rather than to real homogeneity.
We discuss in 
a more quantitative way 
 the behavior in ML surveys in Sec.6, while in Sec.5
we perform a test to proof that the exponent $\alpha \approx 0.4$
is the  real statistical property of the galaxy distribution.

\section{A test for the finite size effects: 
average $N(<m)$}

To prove that the behaviour found in Fig.\ref{f9}, and then the exponent 
$\alpha \approx 0.6$, is connected to large 
fluctuations due to finite size effects 
in the space distribution and not 
for real homogeneity, we have done the following test.
We have adopted the same procedure used for the computation
of the correlation function (\cite{cp92}, \cite{slmp95}), i.e. we make 
an average for $N(<m)$
from all the points of the sample rather than
counting it from the origin only. 

To this aim we have considered a VL sample with $N$ galaxies and we
have built $N$ independent flux-limited surveys in the following way.
We consider each galaxy in the sample as the observer, and 
for each observer we have computed the apparent magnitude of all the 
other galaxies. To avoid any selection effect we consider only the galaxies
 that lie inside a well defined volume around the observer.
This volume is defined by the maximum sphere fully contained in the 
sample volume with the observer as a center.

Moreover we have another selection effect due to the fact that 
our VL sample has been built from a ML survey done with respect 
to the origin. To avoid this incompleteness we have assigned 
to each galaxy 
a constant magnitude $M$. In fact, our aim is to show that 
the inhomogeneity in the space distribution plays the fundamental role 
that determines the shape of the $N(<m)$ relation, 
and the functional form of the luminosity function 
enters in Eq.\ref{q4c} only as an overall normalizing factor.

Once we have computed $N_i(<m)$ from all the points $i=1,..,N$ we then 
compute 
the average. We show in Fig.\ref{f13}
\bef
\vspace{5cm}
\caption{\label{f13}
The average $N(<m)$ in the VL sample VL60.
 The squares crosses refer to  the average $N(<m)$
computed assigning to all the galaxies the same absolute magnitude
$M_{0}=M^*$. The reference line has a slope   $\alpha=0.4$
}
\eef
 and Fig.\ref{f14}
\bef
\vspace{5cm}
\caption{\label{f14}
The average $N(<m)$ in the VL sample VL110.
The squares crosses refer to  the average $N(<m)$
computed assigning to all the galaxies the same absolute magnitude
$M_{0}=M^*$.
The reference line has a slope  $\alpha=0.4$
}
\eef
 the results for VL60 and VL110: 
a very well defined exponent
$\alpha=D/5\approx 0.4$ 
is found in both cases.
This is in fully agreement with the 
average space density (the conditional average density $\Gamma(r)$)
that shows $D \approx 2$ in these VL samples.

We have also performed various tests on 
artificial distributions with a
 priori assigned properties. 
Using the random $\beta-$model algorithm 
\cite{pb88}
and  we have performed  the analysis by 
assigning  to each point of the system the same absolute magnitude. 
The results are in complete agreement with the previous findings:
if we do not perform any average
 the exponent of the GNC is strongly 
affected by the presence of
fluctuations due to finite size effects and we obtain $\alpha \approx 0.6$, 
while if we compute
 the GNC by making an average 
over all the points of the structure 
we find again that the relation $\alpha = D/5$ holds
in a very well approximation
(Fig.\ref{f15} and Fig.\ref{f16}). 
\bef
\vspace{5cm}
\caption{\label{f15}
$(a)$  The space density computed from the vertex 
in  an artificial 
fractal sample with $D=2.5$.
$(b)$ The $N(<m)$ relation for the same sample 
 The reference line has a slope
$\alpha =0.6$.    
}
\eef
\bef
\vspace{5cm}
\caption{\label{f16}
 $(a)$    The average space density computed   
in the  artificial fractal 
sample with $D=2.5$ of Fig.15.
 The reference line has a slope $D-3=-1.$
$(b)$ The average $N(<m)$ for the same 
sample. The reference line has a slope $\alpha = D/5 =0.5.$}
\eef
On the contrary in the homogeneous case 
(Fig.\ref{f12}) one does need to perform any
average to recover the correct scaling properties of 
$N(<m)$, because the system reaches very soon ($r \gtapprox \ell_v$)
the correct scaling properties.

\section{GNC in Magnitude limited catalogs}

We are now able to clarify the problem of ML catalogs.
Suppose to have a certain survey
characterized by a solid angle $\Omega$  and we ask the
following 
question: up to which apparent magnitude limit  $m_{lim}$
we have to push  
 our observations to obtain that 
the majority of the galaxies lie in the statistically significant 
 region ($r \gtapprox  \lambda$)
defined by Eq.\ref{new11} 
Beyond this value of $m_{lim}$ we should recover the genuine
properties of the sample because,
as we have enough statistics, the 
finite size effects self-average. 
From the previous condition for each$\Omega$ we can find a
solid angle $m_{lim}$  so that finally we are able to
obtain $m_{lim}=m_{lim}(\Omega)$ in the following way.

We assume 
a Schecther luminosity function (Eq.\ref{q5f}) and that the fractal dimension
is $D=2$ (we have tested that the final result depends
very weakly on some reasonable values for 
the three parameters used: $D$ for the space
distribution and $\delta$ and $M^*$ for the luminosity function).
As shown in Fig.\ref{f17} 
\bef
\vspace{5cm}
\caption{\label{f17}
An ideal  survey in the $d-M$ space.
If the majority of galaxies lie in the statistically significant region 
then we can obtain a meaningful statistical information from the survey.
This condition  implies that the integral $I_2$ 
is greater than $I_1$.
}
\eef
the requested condition happens when the area $I_2 >I_1$.
We have evaluated numerically these integrals and in Fig.\ref{f18}
\bef
\vspace{5cm}
\caption{\label{f18}
If a survey defined by  the apparent magnitude limit $m_{lim}$
and the solid angle $\Omega$ lie in the  statistically significant
 region
it is possible to obtain the self-averaging properties
of the distribution also with the integral from the vertex.
Otherwise one needs a redshift survey that contains the three
dimensional information, and then one can perform average.
Only in this way it is possible to smooth out the finite size effects.
}
\eef
we show the result that satisfies the requested conditions.
 From the previous figure it follows
that for $m > 19$ the statistically significant region  is reached for
almost {\em any} reasonable 
value of the survey 
solid angle. This implies that 
in the deep surveys, if we have enough statistics,
we   readily  find the right behavior ($\alpha =D/5$)
while it does not happens in a self-averaging way for the 
nearby samples.
Hence the exponent $\alpha \approx 0.4$ 
found in the deep surveys ($m>19$)
is a {\em genuine feature of  galaxy distribution}, and corresponds to 
real correlation properties.
 In the nearby surveys $m < 17$ we do not find 
the scaling region in the ML sample for almost {\em any}
value of the solid angle. 
Correspondingly the value of the 
exponent is subject to the finite size effects,
and to recover the real statistical properties
of the distribution one has to perform an average.

From  the previous discussion it appears now clear why
a change of slope is found at $m \sim 19$: this is
just a reflection of the lower cut-off of the fractal structure
and in the surveys with $m_{lim} > 19$ the self-averaging properties
of the distribution cancel out the finite size effects.
This result depend very weakly on the fractal dimension $D$ and 
on the parameters of the luminosity function $\delta$ and $M^*$
used. Our conclusion is therefore that 
the exponent $\alpha \approx 0.4$ for $m > 19$ 
is a genuine feature of the galaxy distribution and it is related
to a fractal dimension $D \approx 2$, that is found for $m < 19$ in 
redshift surveys only {\em performing averages}.

We note that this result is based 
on the assumption that the Schechter luminosity function
(eq.(15)) holds also at high redshift, or, at least   to $m \sim 20$.
This result is confirmed by the analysis of Vettolani \etal \cite{ve94}
who found that the luminosity function up to $z \sim 0.2$ is in 
excellent agreement with that found in local surveys \cite{dac94}.

Finally we report in Table 2
\begin{table}
\begin{tabular}{lllllll}
\hline
& 	&	 &      &            &     &            \\

Photometric Band         & Driver  & Metcalfe  & Tyson & Lilly &     Djorgovski
 & Jones\\
&	& 	&       &            &        &         \\
\hline
\hline

&	&	&      &            &            &     \\
U      &  -      &	-&	-&	-& 	-&0.5\\
&	&	&      &            &                 &\\
B     & 0.43 & 0.49  & 0.45 & 0.38 & -&-\\
 
&	&	&      &            &           &      \\
V     & 0.39 & -  & - & - & -&-\\
 
&	&	&      &         &   &                 \\
R     & 0.37 & 0.37  & 0.39 & - &-& -\\
 
&	&	&      &            & &                 \\
I     & 0.34 & -  & 0.34 & 0.32 & -&-\\
 
&	&	&      &            &    &             \\
K     & - & -  & - & - & 0.32 &-\\
&	&	&      &          &  &                 \\
\hline
\end{tabular}
\caption{\label{tabbande} 
The exponents of galaxy counts in different spectral bands
}
\end{table}
the exponents of the galaxy counts in different frequency bands, at 
faint magnitudes. We can see that the exponents is lower than $0.6$
in all the case, and it is in the range $0.3 \div 0.5$, so that 
$D$ is in the range $1.5 \div 2.5$.
These differences  can be 
probably related to the multifractal behavior of the luminous matter
distribution \cite{slp95} or to a poor statistics. Only a three dimensional
analysis allows one to decide between these two possibilities.

\section{Definition of a {\em statistically fair} sample}

In the previous sections we have defined the condition to
select a sample large enough to manifest the 
self-averaging properties of a fractal distribution. We now
study in detail a different but related question. Given a 
sample with a well defined volume, which is the minimum number of 
points that it should contain in order to
have a {\em statistically  fair} sample 
even if one computes averages over all the points.

Consider a sample which contains a portion of 
a fractal structure with a lower cut-off $B$. Given the 
geometry of the sample (depth $R$ and solid angle $\Omega$) 
we obtain the value $\lambda$ of the 
{\em minimal statistical length}
according to Eq.\ref{new5} and Eq.\ref{new11}.
Now we investigate what happens if we 
eliminate {\em randomly} more and more points from the sample. 
The fractal dimension and the lower cut-off 
({\em minimal statistical length}) will not be affected by this 
depletion process, because they are related only to 
the intrinsic properties of the fractal structure, i.e to $B$.
Instead, the correlation properties
are more and more affected by a statistical noise 
as we cut the points that contribute to the statistics.
This noise is superimposed to the genuine signal
so that $D$ and $\lambda$ are not changed at all, but
the estimation of their values becomes noisy.
 Obviously, given a finite portion of the original system characterized
by a lower cut-off $B$, it will exist a maximum value of the 
number of points that we can eliminate from the structure in order
to conserve the genuine statistical properties of the 
original distribution.

At this point we can characterize the statistical information in 
each VL sample more
quantitatively. Suppose that the sample volume is a portion of a
sphere with  solid angle $\Omega$ and radius  $R_{VL}$, and that 
the number of points inside  this volume, $N_{VL}$, is 
\be
\label{sf5}
N_{VL} = B_{VL} \frac{\Omega}{4\pi} R_{VL}^D.
\ee
where $B_{VL}$ takes into account the luminosity selection effect.
The original system inside the same volume contains
\be
\label{sf5b}
N = B \frac{\Omega}{4\pi} R_{VL}^D.
\ee

Hence the percentage of galaxies 
present in the sample can be written as
\be
\label{sf6}
p = \frac{N_{VL}}{N} = \frac{B_{VL}}{B}
\ee
In this way we can associate to each VL sample 
a well defined value of $p$ (see Table 3). 
\begin{table}
\caption{The percentage of galaxies $\frac{B_{VL}}{B} \%$ in some
volume limited sample of several surveys. 
$R_{VL}$ is the depth of the volume limited sample
and $N_{VL}$ is the number of points contained.
The all-sky IRAS 
catalogs have a very poor statistics (see text)}
\begin{tabular}{|c|c|c|c|c|}
\hline
	&	&	&	&\\

\rm{Survey} & $\Omega (sr)$ & $R_{VL} (\hmp)$  & $N_{VL}$ & $\frac{B_{VL}}{B} 
\%$ \\
&	&	&	&\\
\hline
\hline
	&	&	&	&     \\
CfA1    & 1.8 	& 40    & 442   & 13 $\%$   \\
	&	&	&	&     \\
CfA1    & 1.8 	& 80 	& 226	& 1.7  $\%$\\
	&	& 	&	&     \\
PP	& 1 	& 60    & 990   & 23  $\%$\\
	&	&	&	&     \\
PP   	& 1 	& 100 	& 688	& 5.7  $\%$\\
	&	& 	&	&     \\
IRAS (2Jy)& 4$\pi$ 	& 40    & 300 & {\em  1.1 $\%$}  \\
	  &	        & 	&	&	\\
IRAS (2Jy)& 4$\pi$ 	& 80    & 250 & {\em 0.2  $\%$} \\
	  &	        &       &	&	\\
IRAS (1.2Jy) & 4$\pi$ 	& 60    & 876   & {\em 1.6  $\%$}  \\
	     &	        &  	&	&	\\
IRAS (1.2Jy) & 4$\pi$ 	& 80    & 766   & {\em 0.8  $\%$}  \\
	     &	        & 	&	&	\\
IRAS (1.2Jy) & 4$\pi$ 	& 100   & 704   & {\em 0.5 $\%$}  \\
	     &		& 	&       &	\\
\hline
\end{tabular}
\end{table}

The crucial point is that the random cut of point must 
stop at a certain limit: beyond this limit one does not have
in the sample enough points to recover the real statistical
properties of the distribution. We
 call {\em statistical fair sample}
a sample that contains a number of points for unit volume 
larger than this limit. The problem is how to define this limit, or, 
in other words, to determine the minimal value of the 
percentage of Eq.\ref{sf6} that allows one to 
recover the genuine information, for example, by the 
two points correlation analysis.
We can proceed in two independent way. The first is by 
analyzing the correlation properties of the VL samples
of Perseus-Pisces and CfA1, the second is 
by studying artificial distributions.

In the VL limited samples of the Perseus-Pisces survey
we can eliminate randomly points up to reach the fraction
contained in the various IRAS samples. We note that 
the correlation function has a clear power law behaviour up to $30 \hmp$
if the percentage remain larger than 
\be
\label{nn2}
p \ge 1 \div 2 \%
\ee
 then it 
shows a cut off towards homogenization  (Fig.\ref{f19}).
\bef
\vspace{5cm}
\caption{\label{f19}
{\em (a)} The average conditional density $\Gamma(r)$
for VL100 of Perseus-Pisces and the whole ML survey.
The reference line has a slope of $-\gamma=-1.1$.
The percentage of galaxies present in the sample is 
 $\sim 6 \%$.
{\em(b)} The same of {\em (a)}
but the  percentage of galaxies present in the sample is 
 $\sim  1.2 \%$.
This is the limiting case 
to recover the statistical properties of the sample,
according to the condition of 
Eq.37
{\em (c)} The same of {\em (a), (b)}
but the percentage of galaxies present in the sample is 
 $\sim  0.5 \%$.
In this case we are well below of the condition of Eq.37.
We can see that at small scale there is a residual power law 
behavior, while at large scale the power law behaviour is broken.
}
\eef
 This is clearly 
spurious and due to the fact that we have reached the 
limit of statistical validity or fairness
of the sample. The apparent homogeneous 
behaviour is related to the fact that in a given sample, 
the mean separation among galaxies $\lambda_v$ grows when the 
number of points decreases ($\lambda_v \sim (V/N)^{1/3}$),
 and when it becomes of the 
same order of the largest void present in the sample volume,
 the 
correlation properties are destroyed. 
This means that the artificial noise introduced by the 
depletion of points, have erased the intrinsical fluctuations of 
the original system. In this
situation see the system as an homogeneous one.

We have considered also an artificial catalog with
a priori assigned properties
generated with the random $\beta-$model algorithm \cite{bp84}.
We show the results in 
Fig.\ref{f20}: 
\bef
\vspace{5cm}
\caption{\label{f20}
{\em (a)} The average conditional density $\Gamma(r)$
for an artificial fractal sample with dimension $D=2$. 
The reference line has a slope of $-\gamma=-1$.
The percentage of galaxies present in the sample is 
 $\sim  100 \%$ (diamonds), 
 $\sim 5 \%$ (crosses) 
(this is the limiting case 
to recover the statistical properties of the sample,
according to the condition of 
Eq.37)
 $\sim  1 \%$ (squares)
In this case we are well
 below of the condition of Eq.37.
We can see that at small scale there is a residual power law 
behavior, while at large scale the power law behaviour is broken.
}
\eef
also in this case we can recover the 
right statistical properties only in the limit of Eq.\ref{nn2}.
Clearly Eq.\ref{nn2} depends from the morphological features of the 
realization of the fractal structure and the percentage can weakly 
fluctuate from a realization to another 

It is possible to compute the $B_{S}$ in the VL samples
of several surveys and we find that in CfA1, PP, ESP 
and LEDA  Eq.\ref{nn2} is 
well satisfied in almost all the VL samples (see Table II). 
On the contrary the VL samples extracted from the 
IRAS $2 Jy$ survey \cite{str90} \cite{str92} and the IRAS $1.2 Jy$ \cite{fis94}
survey
are well beyond the limit of Eq.\ref{nn2} (Table II): 
this is 
the reason why \cite{str90} and \cite{fis94} find that in these 
catalogs $r_0$ does not scale 
with sample depth
as it should be for the fractal case. 
 In fact, in  these cases one observes a
constant density, as the power law correlation 
have been destroyed by a very poor sampling
beyond a certain scale ($ \sim 15 \hmp$) (Fig.\ref{f21}).
\bef
\vspace{5cm}
\caption{\label{f21}
 The average conditional density for some VL samples
of the IRAS 1.2 Jy redshift survey. The reference line has a slope 
of $D \sim 2$ ($\gamma =1$). The crossover towards homogenization is spurious
and due to a poor sampling (see Table 3).
}
\eef
On the other hand Strauss {\em et al.},  
\cite{str90} stress that IRAS galaxies 
belong to the same structures as the optical ones,
and in particular, 
they point out that the infrared galaxies do not fill the voids that 
define the same highly irregular patterns 
seen in the optical samples.
However \cite{str90}  and \cite{fis94} conclude that the IRAS 
galaxies seem to be less correlated than the optical ones,
as the value of the so-called correlation length $r_0$ is smaller than 
that of CfA1. Moreover they claim that the sample is homogeneous and that 
their analysis disprove that galaxy distribution is fractal, at least 
for the infrared galaxies.

On the contrary our results imply that the correlation properties 
as the IRAS galaxies are the same of the optical ones,
even if in the infrared catalogs it is not possible
to recover the correct 
statistical features because of a very poor statistics (Table II).

\section{Experimental implications}

In Sect. 5 we have defined the {\em minimal statistical
 length}
and in Sec.7 we have defined what is a {\em statistically fair sample}.
We consider now
the implications
of these considerations 
 for the optimization of real redshift surveys. 
In fact we are able to establish a quantitative criterion that defines
the statistical validity of a survey, so that it is possible 
to optimize its geometry in order to obtain the maximum 
reliable information.
Eq.\ref{v3} defines the scaling region of a certain survey 
with solid angle $\Omega$
and Eq.\ref{sf6} Eq.\ref{nn2} 
give the conditions for its statistical validity.

Considering a ML sample, 
the survey becomes statistically meaningful, if its 
depth, defined by the 
apparent magnitude limit, and its solid angle  
satisfy  the condition that the majority of galaxies 
lie in the scaling region. In Fig.\ref{f18} 
we have identified the sample fairness 
condition for angular surveys. 
Only the deep surveys 
satisfy this condition, 
while the nearby samples
are all affected by finite size effects. This means that 
one cannot detect statistically meaningful properties from
such surveys. 
Only in the case in which one also measure the redshifts,
as for CfA1 
 for example, one can barely 
recover the genuine properties of the (3d) distribution
as the {\em minimal statistical length} $\lambda$ is of the order of
 the depth of the catalog.
In the case of Perseus-Pisces and CfA2 
one can find some VL samples, 
in which the effective depth $R_{s}$ is appreciably 
larger than $\lambda$. In these cases 
it is possible to study the integral from the vertex
(i.e. $N(<r)$ or $n(r)$) and this allows
 us to extend the analysis up to the deeper
depth observed ($R_s \sim 130 Mpc$). 
But for these same catalogs the ML properties 
are strongly affected by finite size effects because 
most galaxies are to close.

We report in Table 1 the {\em minimal statistical lengths}
for several redshift surveys \cite{hg88}, \cite{pb88}, \cite{gh89}
-\cite{ve94}, \cite{lv92}-\cite{sau90}, 
some of which are published, while others are
 in progress, and this provides a precise 
prediction that can be tested in these surveys.
We have tested such a condition in Perseus-Pisces, CfA1, LEDA and
 ESP 
and the agreement is very good. 

Related to the definition of a redshift catalog we consider the following
problem: given the solid angle of the survey 
(and hence the {\em minimal statistical length}  $\lambda$) 
the crucial point is
that the sample should extend enough beyond $\lambda$, 
having a significant statistics in agreement with Eq.\ref{nn2}.
In order to optimize 
a {\em redshift survey} we can
define a quantitative criterion to select the optimum geometry 
so that 
the finite size effects are minimized. 
A redshift survey is characterized by two parameters:
the solid angle $\Omega$ and the apparent magnitude limit $m_{lim}$.

In order to study the statistical properties of the galaxy
space distribution we have to select from the whole ML survey
a certain Volume Limited (VL) subsample. The number density 
of galaxies ($n_{VL}$)
in such  a sample is given by Eq.\ref{new6} and Eq.\ref{new7}. 
In particular this depends on 
the absolute magnitude limit of such a sample 
which is defined 
by Eq.\ref{q3b}, and is given by (Sec.2)
\be
\label{q3b2}
M_{lim}=m_{lim}-5 \log_{10}R_{VL} -25.
\ee
Hence $n_{VL}$ directly depends on $m_{lim}, \Omega$ and $R_{VL}$, 
where $R_{VL}$ is the maximum depth of the VL sample.
In particular, the {\em first condition} is that $n_{VL}$, which 
refers to 
the {\em deepest VL sample} obtainable from the ML survey,
is such that 
\be
\label{nb}
\frac{B_{VL}}{B} \gtapprox  (2 \div 3) \%
\ee
according to Eq.\ref{nn2}, so that the 
sample is {\em statistically fair}.
The {\em second condition} is that 
\be
\label{rs3}
R_{VL} \gtapprox (2 \div 4) \lambda
\ee
In this way the VL extends more than 
 the {\em minimal statistical length} so that 
the statistical properties can be recovered.
These conditions can be enough to determine the best values of 
$(\Omega,m_{lim}$) but, in order to limit
 the number of redshifts 
measured we can impose also that
 the total number of 
galaxies in the whole Magnitude Limited (ML) catalog 
is at least of the order of 
${\cal N} = (2 \div 3) 10^3$ (this number is a reasonable
choose considering the real experimental surveys). Hence 
from Eq.\ref{q3} and Eq.\ref{q3a} we have ({\it third condition})
\be
\label{q33}
N(<m_{lim}) = B \frac{\Omega}{4 \pi} 
\Phi(\infty) 10^{\frac{D}{5}m_{lim}} \gtapprox {\cal N}
\ee

It is possible to find numerically 
the best solution that 
satisfies the three previous conditions numerically.
In Fig.\ref{f22} 
\bef
\vspace{5cm}
\caption{\label{f22}
The optimum match between the solid 
angle $\Omega$ and the apparent magnitude limit $m_{lim}$,
of a redshift survey 
that satisfies the three conditions explained in the text.
For any of these surveys the number of galaxies in the ML catalog
is constant (${\cal N} \sim (3 \div 5) 10^3$) as well as it is 
constant the number of galaxies the deepest VL sample. 
The depth of such VL 
sample is the order  $\sim (2 \div 4) \lambda$.
}
\eef
we show our results in the $(m_{lim},  \Omega)$
space: for a given $m_{lim}$ the straight line represents
the best solid angle $\Omega$ that maximizes 
the three conditions. The solid angle $\Omega$
decreases with the increasing magnitude as
\be
\Omega \sim 10^{-0.4 m_{lim}}
\ee
In Fig.\ref{f23} 
\bef
\vspace{5cm}
\caption{\label{f23}
 The depth of the deepest VL sample 
as a function of 
the apparent magnitude (see Fig.17)
}
\eef
we show the 
correspondent depth of the deepest VL sample;
$R_{VL}$ grows as
\be
R_{VL} \sim 10^{0.2 m_{lim}}
\ee
We can use these results in the following way:
suppose that our aim is to study the space distribution of galaxies
up to, say $\sim 800 \hmp$.
From Fig.\ref{f23} we can find the $m_{lim}$ of the redshift survey
corresponding to such a distance and from Fig.\ref{f22} the 
relative solid angle $\Omega$. If we follow these conditions we will
obtain the best statistical information from the minimum number
 of redshifts ($\sim 3000$). 
From Fig.\ref{f22} we can note that the ESP survey 
($m_{lim}=19.4$ and  $\Omega =0.006$ \cite{ve94}) is
 almost {\em optimal}
 in this respect, and it can be used to study the 
large scale space distribution 
from $300 \hmp$ up to $\approx (800 \div 900) \hmp$
(\cite{bslmp94} and \cite{psl94}).

\section{The angular two-point correlation function}

 Dogterom \& Pietronero \cite{dp91} 
(see also \cite{cp92}) studied in detail the 
surprising  and subtle  properties of the angular projection of a 
fractal distribution. They find that the angular projection
produces an artificial crossover towards homogenization	 
with respect to the angular density. This crossover
 is artificial (just due to the projection)
 as
 it does not correspond to any physical features of the 
three dimensional distribution. Moreover they showed that 
there is an explicit dependence of 
the angular two point correlation function 
$\omega(\theta)$ on $\theta_M$
the sample angle: this effect has never been taken into account in the 
discussion of real angular catalogs.
These arguments show that it is very dangerous to make
any definite conclusion just 
 from the knowledge of the angular 
distribution. However, there is a point of the discussion 
that remains still open.
If fact, some authors (\cite{pe93}, \cite{ma90}) 
 claim that one of the most important facts 
that disproves the existence of  fractal 
correlations at 
large scales, is the scaling 
of the amplitude
of the two point angular correlation function (ACF)
with sample depth, in the small angles approximation. 
We can now clarify this puzzling situation.
The ACF is defined as \cite{pe93}
\be
\label{acf1}
\omega(\theta) =\frac{<n(\theta_0)n(\theta_0+ \theta)>}{<n>^2}-1
\ee
Assuming that the fractal correlation are only present at small scales,
i.e. that $\xi(r) = (r_0/r)^{\gamma}$,  
it is possible to show that in the 
small angle approximation ($ \theta << 1$)  
one has for the homogeneous case that \cite{pe93}
\be
\label{acf2}
 \omega(\theta)  \sim  \theta^{1-\gamma} (r_0/D_*)^{\gamma}
\ee
where the depth $D_*$ is 
\be
\label{acf3}
D_*= \left( \frac{L_*}{4 \pi f} \right)^{1/2}.
\ee
$L_*$ is the cut-off of the Schechter luminosity function 
and $f$ is the limiting flux density of the survey. 
In the case of a fractal distribution with 
$D <2$ it is easy to show that 
instead of Eq.\ref{acf2} we have \cite{pe93}
\be
\label{acf4}
\omega(\theta)  \sim \theta^{1-\gamma}   
\ee 
so that the difference between the homogeneous and the fractal case
is that in the first case the amplitude of the ACF depends 
on the sample depth $D_*$ while in the second case it is constant.
Peebles \cite{pe93} claims that, since in the real angular catalogs one 
observes the scaling of the amplitude \cite{ma90} and  \cite{gp77},
this provides an evidence against the fractal behavior.
We show now that this conclusion is not correct
because it does not take into account the finite size effects
in real galaxy surveys, as it is the case of the GNC.
In fact, the amplitude of the ACF is strongly related to the behaviour
of the angular density, i.e. to  $N(<m)$.

Eq. \ref{acf4} is obtained under the 
assumption that the  density for  a fractal 
scales as $r^{-\gamma}$:
this is true for the {\em average conditional density}
 in the case of an ideal fractal distribution, 
if the correct scaling regime
 is reached.
The conditional density computed form a single point 
is instead  strongly affected by finite
size effects up to the characteristic 
{\em minimal statistical length} $\lambda$.

In order to illustrate this point 
we present the analysis of the ACF for the Perseus-Pisces 
redshift survey \cite{hg88}. 
 In Fig.\ref{f24} 
\bef
\vspace{5cm}
\caption{
\label{f24}
 The angular correlation function for 
Peruses Pisces limited at $m_{lim} 14.,5$ ($m145i$)
and $15.5$ ($m155i$).
The points refers to different samples of the APM
catalog scaled to the ACF of the Lick survey (Maddox \etal, 1990).
The scaling with depth of the amplitude of the angular correlation function
in this case is due to a finite size effects and 
it is not a proof of homogeneity in space, as we know from the space analysis
that this sample has fractal behavior up to its deeper depth.
}
\eef
we show the behavior of the ACF $\omega (\theta)$ 
for the whole ML survey. We can 
see that there is a clear scaling of the amplitude with the 
apparent magnitude limit of the survey  ($m_{lim}=14.5, 15.5$ 
respectively). We know from the space analysis 
that the galaxy distribution is fractal in this sample and 
therefore this trend is not a consequence of homogeneity, but only of the 
finite size effects 
 that are especially large for the counting from the 
vertex.
In summary the apparent homogeneity inferred
from the angular catalogs has the same origin as the 
exponent $\alpha \approx 0.6$ of the 
galaxy counts at bright magnitudes (small scales).
Both arise from finite size effects
(for a more detailed study on the ACF we refer the reader to 
\cite{ams95}) and do not correspond to the 
real statistical properties.

We note that the exponent of $\omega (\theta)$ is also 
affected by the projection and somewhat higher 
than $1-\gamma \approx 0$, corresponding to $D \approx 2$
obtained in the $3d$ catalog.
 We find in fact that for $\theta << 1$ 
\be
\label{pp1}
\omega (\theta) \sim \theta^{-0.7}
\ee
as it is confirmed in various angular surveys \cite{ma90}  and  \cite{gp77}.
The  approximation of 
the Limber equation for small angles, that gives 
the exponent of the ACF is just $1-\gamma$,
does not hold exactly, may be because we are close to 
the limiting case
$D \approx 2$.

Finally we stress that Park {\em et al.}, 
\cite{pv94}, analyzing the CfA2 redshift surveys,
found the space two point correlation function has an index
 $\gamma \approx 1$
 corresponding to
 $D \approx 2$. The CfA2 surveys is just the three dimensional
counterpart of the Zwicky angular catalog for which Groth \& 
Peebles \cite{gp77} found that the ACF, at small angles,
 scales as 
$\omega(\theta) \sim \theta^{-0.7}$. 
This again shows the difficulties of 
recovering the genuine physical properties from the 
angular analysis alone.

\section{Other astrophysical data}

In Observational Astrophysics there are a lot 
of data that are only angular ones, as the measurements of distances 
is general a very complex task. We briefly discuss here 
the distribution of radiogalaxies, quasars and the $\gamma$-ray burst 
distribution. Our conclusion will 
be that all these data are compatible with a fractal structure with $D \approx 1.6 \div 1.8$
similar to that of galaxies. 

The majority of catalogued radio sources are extragalactic and that some of the strongest
are at cosmological distances. One of the most important information
on radio galaxies distribution has been obtained from the sources counts as 
a function of the apparent flux \cite{con84}. 
Extensive surveys of sources have been 
made at various frequencies in the range  $  0.4 \div 5 \; GHz$. 
In fig.\ref{fsource}
\bef \vspace{5cm} \caption{\label{fsource} 
Normalized differential sources counts 
at $\nu = 1.4 GHz$. {\it Abscissa}
log flux density (Jy). {\it Ordinate} log differential number of sources $n(S)$. The 
solid line represents the behaviour of a fractal structure with $D=1.8$. The agreement
is excellent, except in the bright fluxes region.
} 
\eef 
we show a compendium of sources counts at $\nu=1.4 GHz$ \cite{con84}. 
The differential counts is plotted against the apparent flux. We can see
that in the bright flux region there is a deviation from a power law
function, while for four decades the agreement with 
a fractal distribution with  $  D \approx 1.8$ is excellent. 
Such a behaviour has been usually explained in literature 
as an effect of sources evolution. Here we propose that 
the radio galaxies are fractally distributed, as galaxies, with 
almost the same fractal dimension. The deviation at bright fluxes in
this picture is explained as a 
spurious effect due to  the small scale fluctuations, as in the 
case of galaxy counts. In the other frequency bands 
the situation is nearly the same (\cite{con84}). 
The simple picture of a fractal distribution of radio sources is therefore
fully compatible with the experimental situation. 

In the case of Quasars the situation is almost the same. In fact we find that 
at bright magnitudes ($ 14.75 < B < 18.75$) the exponent of the counts 
is $\alpha \approx 0.,88$, while at faint magnitude it is $\alpha \approx 0.3$
\cite{ht90}. Even in this case we can interpret such a behavior as 
due to the fractal distribution in space with  $  D \approx 1.5$.

Finally we comment the distribution of $\gamma$-ray burst (GRB). 
This is a long-standing problem in Astrophysics and after 20 years of intense studies and 
observations it is still mysterious: the origin of GRBs, in our galaxy or from 
the Cosmos, is a matter of debate
\cite{fis95} \cite{bri96} \cite{har95}. 
We argue here that from the present angular 
and intensity data it is possible to show that 
the space distribution of $\gamma$-ray bursts
sources is fully compatible with a fractal structure 
with $D \approx 1.7$. This result clarifies 
the 
statistical analysis of the available data and points out a fundamental aspect of the
$\gamma$-ray bursts sources distribution.

From the angular catalogs recent available \cite{mee95}
we have substantially it is possible to study three statistical quantities. 
The first one is the number versus apparent intensity distribution that shows 
a deviation from the Euclidean behaviour at low fluxes \cite{mee92} \cite{mee95}.
The second is the 
$V/V_{max}$
test \cite{sch88} that again provides an evidence that the spatial distribution of 
sources is not homogenous \cite{mee92} \cite{mee95}.
 Finally the angular distribution is isotropic within the 
statistical limits \cite{mee95} \cite{bri93} 
 and there is not any evidence for an angular correlation 
or a clustering of bursts towards the galactic center or along the galactic plane of bursts
\cite{mee95} \cite{now94} \cite{qua93} \cite{mee95b}.
These results together indicate that the bursts sources are distributed isotropic but not
homogeneously \cite{mee95}. 
We argue here that these three evidences are fully 
compatible with a fractal distribution of sources with $D \approx 1.7$. 

The first observational fact is the number of burst as a 
function of the apparent flux $N(>f)$.
At bright apparent flux, that are associated to small distance of the sources, one
sees an exponent $-3/2$, that seems to  be in agreement with the homogeneous  case.
This is just a spurious effect that aries form the fact 
this quantity is computed without 
performing any average.
 At faint apparent fluxed, one is integrating the density in the correct
scaling regime, and in this region the genuine statistical properties of the system
can be detected. From the $N(>f)$ relation in the limit of low $f$ we 
can estimate a fractal dimension $D \approx 1.7 \pm 0.1$.

An equivalent test on the homogeneity versus fractal properties is given by the 
$V/V_{max}$ distribution \cite{sch88}, where
\be
\frac{V}{V_{max}} = \left( \frac{C}{C_{lim}} \right)^{-\frac{3}{2}}
\ee
In this ratio $V$ is the volume contained in a sphere extending to the location of the 
 bursts and $V_{max}$ is the volume 
of the sphere extending to the maximum
 distance at which the same burst would be 
detectable by the instrument, whose limiting flux is $C_{lim}$. 
It is simple to show that if the spatial distribution of sources
is described a fractal structure, then we have for the average
\be
<\frac{V}{V_{max}}> = \frac{D}{D+3}
\ee
Even in this case the data show \cite{mee95} that
 $< \frac{V}{V_{max}}> = 0.33 \pm 0.01$
that in terms of fractal dimension means $D \approx 1.5 \pm 0.1$ 
This value is somewhat higher than the fractal dimension 
estimated with the $N(>f)$. This difference is probably due to the fact 
that this test is integrated while the $N(>f)$ is a differential one.

Let us came to the third evidence, i.e. a substantially
 isotropic angular distribution
and a lack of any correlation at small angles. 
As we have seen in the previous section, the projection of
a fractal distribution on the unit sphere conserves the correlations
properties only in the small angles approximation, while 
at large angular scale the long range correlation are destroyed by 
projection effects. 
The angular 
correlation function $\omega(\theta)$ 
has a power law behaviour in the small angle
approximation ($\theta < 10^{\circ}$).  In the available sample \cite{mee95}
the number of points is $N=1122$ distributed over the whole sky. This 
means that at angular distance smaller than $\sim 20^{\circ}$ one does not
see any other object in average, and therefore it is not 
possible to study the angular correlation function at such low
angular separation. 

These results indicate together that the $\gamma$-ray bursts 
Sources are fractally distributed
in space with $D \approx 1.7$. This is value is very similar to that the 
fractal dimension of the galaxy distribution is space  that is
$D \approx 2$ up to some hundreds megaparsec. This "coincidence" can be seen 
as an indication that the 
$\gamma$-ray bursts Sources are associate to a population 
of objects distributed as the visible galaxies.
We stress that a larger sample of bursts will allow one a better determination 
of
the fractal dimension and, if the number of objects for steradians will became
larger, it will be possible to study also the angular correlations in the 
small angle approximation.

\section{Discussion and conclusions}
Since the Hubble's pioneering  work \cite{hu26} 
the galaxy number counts (GNC) has been considered 
as a classical cosmological test and 
one of the most powerful methods to study the galaxy space distribution 
as well as the effect of luminosity and space time evolution 
in different spectral regions \cite{pe93}-\cite{mg94}.

Moreover the apparent exponent $\alpha \approx 0.6$
 at small scale appears to be 
in contradiction with the fractal distribution derived 
from the full correlation analysis in the redshift catalogs.
However the discussion of the behaviour of the GNC 
neglects the effect of the 
non-analyticity of fractal distributions.
A fractal distribution is characterized by having fluctuations 
at all scales and, 
in dealing  with real fractals
 characterized by a lower cut-off, 
a crucial point is the effect of finite size 
fluctuations.
The lower cut-off can be 
defined by the Voronoi volume for 
a Poisson distribution. 
In case of galaxy surveys
the Voronoi distance represents the minimal 
average distance between galaxies, and 
the condition that the survey volume
should contain a large number of Voronoi volumes, 
defines  the {\em minimal statistical length} 
$\lambda$ beyond which the statistical properties 
are detectable. 

A Poisson distribution is characterized by 
small amplitude fluctuations that 
represent small perturbations of 
a well defined average. A fractal distribution
is characterized by fluctuations that can be as 
large as the system 
itself, so that the concept of average density 
looses its physical meaning.
These fluctuations define all the statistical properties
of the distribution.
The qualitative difference between
 the fluctuations in these
two cases is crucial to understand the 
importance of finite size effects.
While for a homogeneous distribution one readily
recovers the genuine statistical properties of
the sample at distance, say, three times
the Voronoi length, in the fractal case
 it is necessary to have 
 a linear size of at least
 ten times that of the  Voronoi volume.
For averages ($\Gamma(r), \Gamma^{*} (r)$) 
one recovers the correct properties at appreciably 
smaller length. 

The main problem of the standard analysis of the GNC
is related to the fact that one computes $N(<m)$ only from 
one point, the origin. In this situation the 
fluctuations  due to the finite size of the system 
dramatically affect the behaviour of the GNC
to rather large scales.

With the aim of clarifying this point,
we have studied
 in detail the behaviour of the GNC in a redshift survey,
the Perseus-Pisces,
where it is possible to control the behaviour of the 
real space density.
We find  that the space distribution is 
well described by a fractal with dimension $D \approx 2$
up to $130 h^{-1}Mpc$ (\cite{slmp95}).
From the GNC analysis, we 
 find that the exponent 
$\alpha \approx 0.6$ is instead related to the 
presence of strong finite size 
fluctuations in the density $n(r)$. 
These strong fluctuations affect the behaviour not only 
of $n(r)$ but also of $N(<m)$ 
and make the exponent of the GNC similar 
to the case of a homogeneous spatial distribution, i.e. 
$\alpha \approx 0.6$. This is because a very fluctuating 
behavior in the density 
can be roughly approximated 
by a constant density. In fact 
at small distances ($r << \lambda$) 
one finds almost no galaxies. Then the number 
of galaxies start to grow but this regime
is strongly affected by fluctuations. Finally the correct scaling 
region is reached for $r \approx \lambda$, and the
regime of raise and fall with strong fluctuations
can be roughly approximated to a constant 
one.
To show that this is indeed the case, 
we consider in detail 
the GNC is some VL samples 
of the Perseus-Pisces survey.
In the deeper VL samples, where the spatial extension of the 
survey  allows the self-averaging of 
 the finite size fluctuations ($ r > \lambda$) 
and it 
allows one to recover the smooth 
power law behavior for the density,
the exponent of the GNC is simply $\alpha = D/5 \approx 0.4$.
If instead one computes the
 density only from one,
 the exponent is $\alpha \approx 0.6$ 
because of the finite size 
fluctuations. This happens also
in the whole magnitude limit survey.
The crucial point is  that the exponent 
is $\alpha \approx 0.6$ in this case is not associated to  homogeneity
in space, but it is due only to the presence of finite size fluctuations 
 in the space distribution that do not self-average because the $N(<m)$ relation
is computed only from one point, the origin.
On the contrary if we perform 
 an average from all the points
of the survey for $N(<m)$ 
we can readily recover  the same exponent
found with the spatial analysis   ($\alpha = D/5 \approx 2$).
 We have tested these results
in some artificial catalogs with a priori assigned properties 
(fractal and homogeneous), finding 
clear confirmations for such a behavior.

Having this result in mind we can clarify now 
the problem of the GNC behaviour. In deeper 
optical samples it 
is found that $\alpha\approx 0.4$. Now we can interpret this result
as the 
genuine one of a large
 enough sample where, as in the case of the deeper
VL samples, the finite size effects are averaged 
out for the presence of structures and voids in such a manner that 
the scaling region is reached
and the self-averaging properties of the fractal structure are present. 
 This situation can be described in a more quantitative way.
We have introduced in 
 Section 4 the {\em minimal statistical length} $\lambda$
that defines the 
length beyond which 
this behavior occurs. 
In particular this length is directly related to {\em the 
lower cut-off of the galaxy distribution and to the solid
angle of the survey}.
In Section 5 we have considered in detail the
 change of slope
in ML surveys in the $B$-band, that are the most 
significant from a statistical point of view.

 The crucial point that arises from our analysis is that
for $ m >19$ the ML survey extends enough in the statistically significant region,
i.e. $r > \lambda$, 
for almost 
all the reasonable values of the solid angle 
of the surveys, so that  the 
genuine behavior can be recovered also in this case, because the
the finite size fluctuations due to the lower cut-off are averaged out.
So we can relate the exponent found in the deep galaxy 
catalogs ($m > 19$) 
with what we find at small magnitude and redshift 
in the appropriate way,
concluding that $\alpha\approx 0.4$ is connected to the 
fractal distribution in space with dimension $D \approx 2$ 
[58] that we observe at small redshifts and magnitudes
 (Fig.\ref{f1}).

Moreover all the different values of the exponent of the GNC 
$\alpha$ found in different 
spectral regions are   associated with the 
presence of these spatial fluctuations and
 are not intrinsic features of the galaxy luminosity distribution.
For example, in $R$-band the results are quite in agreement 
with those in the $B$-band as the exponent is $\alpha \approx 0.4$ 
in both cases. In the $K-$band we find $\alpha \approx 0.67$
in the range $K <17$ and this behavior
is clearly due to finite size fluctuations only.
For $K>17$ the exponent $\alpha \approx 0.3$, and this can be 
probably related   to the multifractal behavior of the luminous matter
distribution \cite{slp95} or to a poor statistics. Only a three dimensional
analysis allows one to decide between these two
possibilities.

Therefore our main conclusion is that the 
analysis of the GNC shows a clear and
strong evidence that the fractal distribution
seen in 
the correlation analysis 
in local redshift surveys, with dimension $D \approx 2$,
extends to the deepest depths
so that we have a unique fractal structure in the 
magnitude range $12 \ltapprox m \ltapprox 28$ 
with $D \approx 2$,
that is to say the largest scales ever
 probed for luminous matter.

In addition we have shown that the counts of radio galaxies, Quasars
and $\gamma$ ray bursts are fully compatible with a fractal 
distribution in space with dimension $D \approx 1.7$. 
The behavior of the real data for these objects looks like 
the case of galaxy counts.  For radio galaxies and Quasars
we can conclude that the genuine scaling behaviour can be 
found only at faint fluxes, and it is related to the 
fractal properties of the space distribution rather than 
to the effects of evolution of space time. In the case 
of $\gamma$ ray bursts our conclusion is that 
there is a strong evidence that the sources
are associate to a population 
of objects distributed as the visible galaxies, as they have 
almost the same fractal dimension.

The 
{\em minimal statistical length} , that is related to the fractal
lower cut-offs and the solid angle of a survey  
has important consequences also from an experimental point of view.
In fact, only 
beyond such a length one  can 
find the correct scaling 
properties.
This gives a quantitative criterion 
to define the statistical validity of a survey and for the optimization
of its geometry 
in order to 
to derive the maximum reliable information. 
On this basis we can {\em predict} the 
expected statistical properties of several angular and redshift
surveys (Table 1, Fig.\ref{f18} and Fig.\ref{f22}). Moreover from the 
knowledge of the lower cut-off, we can 
establish a quantitative condition to define a {\em statistically
fair sample}, i.e. a sample that is large enough to 
allow one to recover the genuine statistical properties of the 
distribution. Such a condition is related to the 
number of points present in the sample to its space 
extension, and it depends explicitly on  the lower cut-off 
of the fractal structure.

Finally have we considered the consequences of the finite 
size effects on the determination of the amplitude of the 
angular two point
correlation function. Our analysis implies that this quantity is
strongly affected by the presence of finite size effects,
as it is determined from a single point, so that 
it suffers of the same problems of the GNC.
Indeed the amplitude of the GNC is directly related 
to the angular density, i.e. to $N(<m)$.
As a consequence of these spurious effects 
one has the impression of an apparent
homogenization shown 
by the scaling with depth 
of the amplitude of the ACF, in 
the small angles approximation.
This scaling has been considered \cite{pe93} 
as an 
evidence for an homogeneous distribution of 
galaxies, while our results show that it arises only from
finite size fluctuations 
 due to the presence of a lower cut-off
in the galaxy distribution. This analysis 
shows again that the analysis of the angular properties
for distribution characterized by long-range correlations 
is very subtle and requires a careful 
treatment.

The result that the galaxy distribution can be described
 as fractal
structure with $D \approx 2$ from small scales, up
to the deepest scale ever probed for visible matter 
has dramatic consequences for the standard scenario \cite{bslmp94}.
At relatively small scales
the evolutionary effects are certainly negligible.
However the observation of the same behavior in the 
entire range $12 \ltapprox  m \ltapprox 28$ 
raises a puzzling problem for the 
standard framework because one would
expect at large scales some modifications
due to the space-time evolution effects and 
the galaxy evolution mechanism.

\section*{Acknowledgments}
We are very pleasant to thank Dr. M. Haynes and 
Dr. R. Giovanelli for having given us the possibility
of analyzing the Perseus-Pisces redshift survey.
We thank also  Dr. Yuri Baryshev  for useful comments and 
suggestions 
and Dr. Luca Amendola for continuous discussions, suggestions and 
for his help in the angular analysis.
Finally we thank G. Paturel and H. Di Nella 
for stimulating discussions on the properties 
of the galaxy counts in the LEDA sample.

\end{document}